\documentclass[aoas,nameyear,dvips]{arximspdf}

\doi{10.1214/10-AOAS419}
\volume{5}
\issue{2A}
\pubyear{2011}
\firstpage{824}
\lastpage{842}



\begin{document}
\begin{frontmatter}

\title{Variance estimation for nearest neighbor imputation for US
Census long form data\thanksref{T1}}

\thankstext{T1}{This paper is released to inform interested parties
of ongoing research and to encourage discussion. The views expressed on
statistical, methodological, technical, or operational issues are those
of the authors and not necessarily those of the US Census Bureau.}

\runtitle{Nearest neighbor imputation}

\begin{aug}
\author[A]{\fnms{Jae Kwang} \snm{Kim}\corref{}\ead[label=e1]{jkim@iastate.edu}},
\author[A]{\fnms{Wayne A.} \snm{Fuller}\ead[label=e2]{waf@iastate.edu}}
\and
\author[B]{\fnms{William R.} \snm{Bell}\ead[label=e3]{william.r.bell@census.gov}}

\runauthor{J. K. Kim, W. A. Fuller and W. R. Bell}

\affiliation{Iowa State University, Iowa State University and US
Bureau of Census}

\address[A]{J. K. Kim\\
W. A. Fuller\\
Department of Statistics\\
Iowa State University\\
Ames, Iowa 50011\\
USA \\
\printead{e1}\\
\phantom{E-mail:\ }\printead*{e2}} 

\address[B]{W. R. Bell\\
US Bureau of Census\\
Washington DC, 20233\\
USA\\
\printead{e3}}
\end{aug}

\received{\smonth{9} \syear{2009}}
\revised{\smonth{9} \syear{2010}}

%
\begin{abstract}
Variance estimation for estimators of state, county, and school
district quantities derived from the Census 2000 long form are
discussed. The variance estimator must account for (1) uncertainty due
to imputation, and (2)~raking to census population controls.
An imputation procedure that imputes more than one value for each
missing item using donors that are neighbors is described and the
procedure using two nearest neighbors is applied to the Census long
form. The Kim and Fuller [\textit{Biometrika} \textbf{91} (2004) 559--578]
method for variance estimation under fractional hot deck imputation is
adapted for application to the long form data. Numerical results from
the 2000 long form data are presented.
\end{abstract}

%

\end{frontmatter}

\section{Introduction}\label{sec.1}

In Census 2000 income data were collected on the long form that was
distributed to about one of every 6 households in the United States.
These data were used to produce various income and poverty estimates
for the US, and for states, counties, and other small areas. The state
and county income and poverty estimates from the Census 2000 long form
sample have been used in various ways by the Census Bureau's Small Area
Income and Poverty Estimates (SAIPE) program. The poverty estimates
produced by SAIPE have been used by the US Department of Education in
allocating considerable federal funds each year to states and school
districts. In 2008 the Department of Education used SAIPE estimates,
directly and indirectly, to allocate approximately \$16 billion to
school districts.

The Census 2000 long form had questions for eight different types of
income for each individual in a household. (For details, see Table
\ref{table1} in Section~\ref{s5}.) If there was nonresponse for an income
item, a version of nearest neighbor imputation (NNI) was used, where
the nearest neighbor was determined by several factors such as response
pattern, number of household members, and other demographic
characteristics. NNI is a type of hot deck imputation that selects the
respondent closest, in some metric, to the nonrespondent, and inserts
the respondent value for the missing item. Most imputation rates for
income items in the Census 2000 long form data were more than double
the corresponding imputation rates from the 1990 census [Schneider
(\citeyear{s2004}), pages 17--18, and Table 1, page 27]. For example,
the Census 2000 imputation rate for wage and salary income was 20\%,
while in 1990 it was 10\%, and for interest and dividend income the
imputation rates were 20.8\% in 2000 and 8.1\% in 1990. Overall, 29.7\%
of long form records in 2000 had at least some income imputed, compared
to 13.4\% in 1990. Given the 2000 imputation rates, it is important
that variance estimates for income and poverty statistics reflect the
uncertainty associated with the imputation of income items.

The Census Bureau performed nearest neighbor imputation for eight
income items in producing the long form estimates. The estimation
procedure had been implemented and the estimates were not subject to
revision. Our task was to estimate the variances of the existing long
form point estimates that are used by the SAIPE program. The problem is
challenging because of the complexity of the estimates. While total
household income is a simple sum of the income items for persons in a
household, and average household income (for states and counties) is a
simple linear function of these quantities, our interest centers on (i)
median household income, and (ii) numbers of persons in poverty for
various age groups. Poverty status is determined by comparing total
family income to the appropriate poverty threshold, with the poverty
status of each person in a family determined by the poverty status of
the family. For such complicated functions of the data, the effects of
imputation on variances are difficult to evaluate.

It is well known that treating the
imputed values as if they are observed and applying a standard variance
formula leads to underestimation of the true variance. Variance
estimation methods accounting for the effect of imputation have been
studied by Rubin (\citeyear{r1987}), Rao and Shao (\citeyear
{rs1992}), Shao and Steel (\citeyear{ss1999}),
and Kim and Fuller (\citeyear{kf2004}), among others. Sande (\citeyear
{s1983}) reviewed the NNI
approach, Rancourt, S{\"{a}}rndal, and Lee (\citeyear{rsl1994})
studied NNI under a
linear regression model, and Fay (\citeyear{f1999}) and Rancourt
(\citeyear{r1999}) considered
variance estimation in some simple situations. Chen and Shao (\citeyear
{ch2000})
gave conditions under which the bias in NNI is small relative to the
standard error and proposed a model-based variance estimator. Chen and
Shao (\citeyear{cs2001}) described a jackknife variance estimator.
Shao and Wang
(\citeyear{sw2008}) discussed interval estimation and Shao (\citeyear
{s2009}) proposed a simple
nonparametric variance estimator.

Our approach to estimating variances under NNI is based on the
fractional imputation approach suggested by Kalton and Kish (\citeyear
{kk1984}) and
studied by Kim and Fuller (\citeyear{kf2004}). In fractional
imputation, multiple
donors, say, $M$, are chosen for each recipient.
We combine fractional imputation with the nearest neighbor criterion of
selecting donors, modifying the variance estimation method described in
Kim and Fuller (\citeyear{kf2004}) to estimate the variance due to
nearest neighbor
imputation.
Replication permits estimation of variances for parameters such as
median household income and the poverty rate. Also, replication is used
to incorporate the effect of raking, another feature of the estimation
from the Census 2000 long form sample.

It should be noted that the official estimation and imputation
procedures for the long form were fixed and production was completed
before the research described here was even started. Hence, our
objective was to develop variance estimates, accounting for imputation
and raking, for the production point estimates, not to explore
alternative imputation procedures in an attempt to improve the point
estimates. Thus, we used $M =2$ nearest neighbor imputations in
developing variance estimates for the production long form estimates
that used $M = 1$ nearest neighbor imputation.

The paper is organized as follows. In Section \ref{s2} the model for
the NNI
method and the properties of the NNI estimator are discussed. In
Section \ref{s3} a variance estimation method for the NNI estimator is
proposed. In Section~\ref{s4} the proposed method is extended to
stratified cluster sampling. In Section~\ref{s5} application of the approach
to the Census 2000 long form income and poverty estimates is described.

\section{Model and estimator properties}\label{s2}

Our finite universe $U$ is the census population of the United States.
The Census Bureau imputation procedure defines a measure of closeness
for individuals. Let a neighborhood of individual $g$ be composed of
individuals that are close to individual $g$, and let~$B_g$ be the set
of indices for the individuals in the neighborhood of individual $g$.
We assume that it is appropriate to approximate the distribution of
elements in the
neighborhood by
\begin{equation}\label{1}
y_j \stackrel{\mathrm{i.i.d.}}{\sim}( \mu_g, \sigma_g^2 ),\qquad j
\in B_g ,
\end{equation}
where $ \stackrel{\mathrm{i.i.d.}}{\sim}$ denotes independently and
identically distributed. Chen and Shao (\citeyear{ch2000}) have given conditions
such that it is possible to define a sequence of samples, populations,
and neighborhoods so that the distribution of $y_i$ can be approximated
by that of (\ref{1}). See also Section B in the supplemental article
[Kim, Fuller, and Bell (\citeyear{kfb2010})] for an alternative justification of
(\ref{1}). These conditions do not necessarily hold for our population
because the neighbors are defined by discrete variables. If response is
independent of $y$ and if the value of the discrete variables are the
same for all elements in $B_g$, then (\ref{1}) holds when the original
observations are independent. We feel (\ref{1}) is reasonable because
the sample is large relative to a neighborhood composed of three sample
individuals. We assume that response is independent of the $y$-values
so that the distribution (\ref{1}) holds for both recipients and
donors.

Let $\hat{\theta}_n$ be an
estimator based on the full sample. We write an estimator that is
linear in $y$ as
\[
\hat{\theta}_n = \sum_{i \in A} w_i y_i ,
\]
where $A$ is the set of indices in the sample and the weight $w_i$
does not depend on~$y_i$. An example is the estimated total
$\hat{T}_y = \sum_{ i \in A} \pi_i^{-1} y_i ,$ where $\pi_i$ is the
selection probability. Let $V( \hat{\theta}_n )$ be the variance of
the full sample estimator. Under model~(\ref{1}) we can write
\[
y_i = \mu_i + e_i ,
\]
where the $e_i$ are independent $( 0, \sigma_i^2 )$ random variables
and $\mu_i$ is the neighborhood mean. Thus, $\mu_i = \mu_g$ and
$\sigma_i^2= \sigma_g^2$ for $i \in B_g$. Then, under model~(\ref{1})
and assuming that the sampling design is ignorable under the model in
the sense of Rubin (\citeyear{r1976}), the variance of a linear estimator of the
total $T_y = \sum_{i \in U} y_i $ can be written
\[
V\biggl\{ \sum_{i \in A} w_i y_i - T_y\biggr\} =V\biggl\{ \sum_{i \in A} w_i \mu_i -
\sum_{i \in U } \mu_i \biggr\} + E\biggl\{ \sum_{i \in A} ( w_i^2 - w_i )
\sigma_i^2 \biggr\}.
\]

Assume that $y$ is missing for some elements and assume there are
always at least $M$ observations on $y$ in the neighborhood of each
missing value, where in the Census long form application, $M=2$.
Let an imputation procedure be used to assign $M$
donors to each recipient. Let $w_{ij}^*$ be the fraction of the
original weight allocated to donor $i$ for recipient $j$, where
$\sum_{i} w_{ij}^* = 1$. If we define
\[
d_{ij}=\cases{
1, &\quad if $y_i$ is used as a donor for $y_j$, \cr
0, &\quad otherwise,
}
\]
then one common choice for $w_{ij}^*$ is $w_{ij}^*= M^{-1} d_{ij} $
for $i \neq j$. Then
\[
\alpha_i = w_i + \sum_{j \neq i} w_j w_{ij}^* = \sum_{j \in A} w_j
w_{ij}^*
\]
is the total weight for donor $i$, where it is understood that
$w_{ii}^*=1$ for a~donor donating to itself. Thus, the imputed linear
estimator is
\[
\hat{\theta}_I = \sum_{j \in A} w_j y_{Ij} = \sum_{ i \in A_R}
\alpha_i
y_i ,
\]
where $A_R$ is the set of indices for the $n_R$ respondents and the
mean imputed value for recipient $j$ is
\begin{equation}\label{7}
y_{Ij} = \sum_{ i \in A} w_{ij}^* y_i .
\end{equation}
Note that $y_{Ij}=y_i$ if $j$ is a respondent. Then, under model
(\ref{1}),
\begin{equation}\label{8}
V( \hat{\theta}_I - T_y ) = V\biggl\{ \sum_{i \in A} w_i \mu_i -\sum_{i
\in
U} \mu_i
\biggr\} + E\biggl\{ \sum_{ i \in A_R } ( \alpha_i^2 - \alpha_i ) \sigma_i^2
\biggr\},
\end{equation}
where $A_R$ is the set of indices of respondents. The variance
expression (\ref{8}) is smaller for larger $M $, $1 \le M \le n_R$, as
long as model (\ref{1}) holds for the $M$ nearest neighbors. See Kim
and Fuller (\citeyear{kf2004}).

\section{Variance estimation}\label{s3}

Let the replication variance estimator for the complete sample be
\begin{equation}\label{9}
\hat{V} ( \hat{\theta} ) = \sum_{k=1}^L c_k \bigl( \hat{\theta}^{(k)} -
\hat{\theta} \bigr)^2 ,
\end{equation}
where $\hat{\theta}$ is the full sample estimator, $\hat{\theta}^{(k)}$
is the $k$th estimate of $\theta_N$ based on the observations included
in the $k$th replicate, $L$ is the number of replicates, and $c_k$ is a
factor associated with replicate $k$ determined by the replication
method. Assume that the variance estimator $\hat{V} ( \hat{\theta} )$
is design unbiased for the sampling variance of $\hat{\theta}$. If the
missing $y_i$ are replaced in (\ref{9}) with $y_{Ij}$ of (\ref{7}),
the resulting variance estimator
$\hat{V}_{\mathrm{naive}} ( \hat{\theta} )$ satisfies
\begin{equation}\label{10}%
E\{ \hat{V}_{\mathrm{naive}} ( \hat{\theta} ) \} = V \biggl\{ \sum_{i \in A} w_i
\mu_i
- \sum_{ i \in U} \mu_i \biggr\} + E \Biggl\{ \sum_{k=1}^L \sum_{i \in A_R }
c_k \bigl(
\alpha_{i1}^{(k)} - \alpha_i \bigr)^2 \sigma_i^2 \Biggr\},\hspace*{-30pt}%
\end{equation}
where $\alpha_{i1}^{(k)} = \sum_{j } w_j^{(k)} w_{ij}^* $ and
$w_j^{(k)}$ is the weight for element $j$ in replicate $k$. The weights
$\alpha_{i1}^{(k)} $ are called the naive replication weights.

We consider a procedure in which the individual $w_{ij}^*$ are modified
for the replicates, with the objective of creating an unbiased variance
estimator.
Let~$w_{ij}^{*(k)}$ be the
replicated fractional weights of unit $j$ assigned to donor $i$ at the
$k$th replication. Letting
\[
\hat{\theta}_I^{(k)} = \sum_{ i \in A_R } \alpha_i^{(k)} y_i ,
\]
where $\alpha_i^{(k)} = w_i^{(k)} + \sum_{j \neq i}w_j^{(k)}
w_{ij}^{*(k)} = \sum_{j \in A} w_j^{(k)} w_{ij}^{*(k)} ,$ define a
variance estimator by
\[
\hat{V} ( \hat{\theta}_I ) = \sum_{k=1}^L c_k \bigl( \hat{\theta
}_I^{(k)} -
\hat{\theta}_I \bigr)^2 .
\]

The expectation of the variance estimator $\hat{V}(
\hat{\theta}_I )$ is
\begin{eqnarray}\label{6b}
E\{ \hat{V} ( \hat{\theta}_I ) \}
&=&
E\Biggl[ \sum_{k=1}^L \biggl\{ \sum_{i \in
A_R}\bigl( \alpha_i^{(k)} -\alpha_i \bigr) \mu_i \biggr\}^2 \Biggr]\nonumber
\\[-8pt]\\[-8pt]
&&{}+
E\Biggl[ \sum_{i \in A_R
} \Biggl\{
\sum_{k=1}^L c_k \bigl( \alpha_{i}^{(k)} - \alpha_i \bigr)^2 \Biggr\} \sigma_i^2 \Biggr].\nonumber
\end{eqnarray}
Because the $w_{ij}^{*(k)}$ satisfy
\begin{equation}\label{con1}
\sum_{ i \in A_R} w_{ij}^{*(k)} = 1
\end{equation}
for all $j$, then, under the model (\ref{1}), ignoring the smaller
order terms,
\begin{eqnarray*}
E\Biggl\{\sum_{k=1}^L\biggl[\sum_{i\in A_R}\bigl(\alpha_i^{(k)}-\alpha_i\bigr)\mu_i\biggr]^2\Biggr\}
&=&
E\Biggl\{\sum_{k=1}^L\biggl[\sum_{i\in A}\bigl(w_i^{(k)}-w_i\bigr)\mu_i\biggr]^2\Biggr\}
\\
&=&
V\biggl(\sum_{i\in A}w_i\mu-\sum_{i\in U}\mu_i\biggr).
\end{eqnarray*}
Thus, the bias of the variance estimator $\hat{V}(
\hat{\theta}_I )$ is
\[
\mathit{Bias} \{\hat{V} ( \hat{\theta}_I ) \}
= E\Biggl\{ \sum_{i \in A_R } \Biggl[ \sum_{k=1}^L c_k \bigl( \alpha_{i}^{(k)} -
\alpha_i \bigr)^2-( \alpha_i^2 -\alpha_i ) \Biggr] \sigma_i^2 \Biggr\}.
\]
If the replicated fractional weights were to satisfy
\begin{equation}\label{con3}
\sum_{k=1}^L c_k \bigl( \alpha_i^{(k)} - \alpha_i \bigr)^2 = \alpha_i^2 -
\alpha_i
\end{equation}
for all $i \in A_R$, then the bias would be zero. However, it is
difficult to define replicate weights that satisfy (\ref{con3}).
Therefore, we consider the requirement
\begin{equation}\label{con2}
\sum_{k=1}^L c_k \biggl\{ \bigl( \alpha_i^{(k)} - \alpha_i \bigr)^2 + \sum_{ t \in
D_{Ri} } \bigl( \alpha_t^{(k)} - \alpha_t \bigr)^2 \biggr\} = \alpha_i^2 - \alpha_i
+ \sum_{t \in D_{Ri} } ( \alpha_t^2 - \alpha_t ),\hspace*{-30pt}
\end{equation}
where $D_{Ri} = \{ t ; \sum_{j \in A_M } d_{ij} d_{tj} = 1, t \neq i
\} $ is the set of donors, other than $i$, to recipients from donor
$i$. Under assumption (\ref{1}), the recipients in the neighborhood of
donor $i$ have common variance and (\ref{con2}) is a sufficient
condition for unbiasedness.

We outline a replication variance estimator that assigns fractional
replicate weights such that
(\ref{con1}) and (\ref{con2}) are satisfied. There
are three types of observations in the data set: (1) respondents that
act as donors for at least one recipient, (2)~respondents that are
never used as donors, and (3) recipients. The naive replicate weights
defined in (\ref{10}) will be used for the last two types. For donors,
the fractional weights $w_{ij}^*$ in replicate $k$ will be modified
to satisfy (\ref{con1}) and (\ref{con2}).

We first consider jackknife replicates formed by deleting a single
element. The next section considers an extension to a grouped jackknife
procedure.
Let the superscript $k$ denote the
replicate where element $k$ is deleted. First the replicates for the
naive variance estimator (\ref{10}) are computed, and the sum of
squares for element~$i$ is computed as
\[
\sum_{k=1}^L c_k \bigl( \alpha_{i1}^{(k)} - \alpha_i \bigr)^2 = \phi_i,\qquad i \in
A_R,
\]
where $\alpha_{i1}^{(k)}$ is defined following (\ref{10}).

In the second step the
fractions for replicates for donors are modified.
Let
the new fractional weight in replicate $k$ for the value donated by $k$
to $j$ be
\begin{equation}\label{12}
w_{kj}^{*(k)} = w_{kj}^*( 1 - b_k) ,
\end{equation}
where $b_k$ is to be determined. Let $t$ be one of the other $M-1$
donors, other than~$k$, that donate to $j$. Then, the new fractional
weight for donor $t$ is
\begin{equation}\label{13}
w_{tj}^{*(k)} = w_{tj}^* + ( M-1 )^{-1} b_k w_{kj}^*.
\end{equation}
For $M=2$ with $w_{kj}^*=w_{tj}^*=0.5$, $w_{kj}^{*(k)} = 0.5(1-b_k)$
and $w_{tj}^{*(k)} = 0.5(1+b_k)$.

For any choice of $b_k$,
condition (\ref{con1}) is satisfied. The variance estimator will be
unbiased if $b_k$ satisfies
\begin{eqnarray}\label{14}
&&
c_k\biggl(\alpha_{k1}^{(k)}-\alpha_k-b_k\sum_{j\in A_M}w_j^{(k)}w_{kj}^*\biggr)^2-c_k \bigl(\alpha_{k1}^{(k)}-\alpha_k \bigr)^2\nonumber
\\
&&\qquad{}+
\sum_{t\in D_{Rk}}c_k\biggl[\alpha_{t1}^{(k)}-\alpha_t+b_k(M-1)^{-1}\sum_{j\in A_M}w_j^{(k)}w_{kj}^*d_{tj}\biggr]^2
\\
&&\qquad{}-
\sum_{t\in D_{Rk}}c_k\bigl(\alpha_{t1}^{(k)}-\alpha_t\bigr)^2=\alpha_k^2-\alpha_k-\phi_k,\nonumber
\end{eqnarray}
where $D_{Rk}$ is defined following (\ref{con2}).
The difference $\alpha_k^2-\alpha_k - \phi_k$ is the
difference between the desired sum of squares for observation $k$ and
the sum of squares for the naive estimator. Under the assumption of a
common variance in a neighborhood and the assumption that the variance
estimator~$\hat{V}(\hat{\theta})$ of~(\ref{9}) is unbiased for the full
sample, the resulting variance estimator with~$w_{ij}^{*(k)}$ defined
by (\ref{12})--(\ref{14}) is unbiased for the imputed sample. An
illustration of the construction of replicates for variance estimation
is provided in Section A of the supplement [Kim, Fuller, and Bell
(\citeyear{kfb2010})].

\section{Extension}\label{s4}

The proposed method in Section \ref{s3} was described under the situation
where the jackknife replicates are formed by deleting a single element.
In practice, grouped jackknife is commonly used where the jackknife
replicates are often created by deleting a group of elements. The group
can be the primary sampling units (PSU) or, as in the Census long form
case, groups are formed to reduce the number of replicates. In the
discussion we use the term PSU to denote the group. To extend the
proposed method, assume that we have a sample composed of PSUs and let
PSU $k$ be deleted to form a replicate. Let $\mathcal{P}_k$ be the
indices of the set of donors in PSU $k$ that donate to a recipient in a
different PSU. For fractional imputation of size~$M$, let the
fractional replication weight in replicate $k$ for the value donated by
element $i$ in PSU $k$ to $j$ be
\begin{equation}
w_{ij}^{*(k)} =
w_{ij}^* ( 1 - b_k )\qquad \mbox{if } i \in
\mathcal{P}_k \mbox{ and } M \neq M_{jk},
\label{15}
\end{equation}
where $b_k$ is to be determined and $M_{jk} = \sum_{ i \in
\mathcal{P}_k } d_{ij} $ is the number of donors to recipient $j$ that
are in PSU $k$. Note that (\ref{15}) is a generalization of (\ref{12}).
The corresponding replication fraction for a donor to a recipient
$j$, where the donor is not in PSU $k$, is
\[
w_{tj}^{*(k)} = w_{tj}^* ( 1+ \Delta_{jk} b_k d_{ij}) \qquad\mbox{for } t
\in\mathcal{P}_k^c \mbox{ and } i \in\mathcal{P}_k ,
\]
where
\[
\Delta_{jk} = \frac{ \sum_{i \in\mathcal{P}_k } w_{ij}^* }{ \sum
_{i \in\mathcal{P}_k^c } w_{ij}^*
}.
\]
The determining equation for $b_k$ is
\begin{eqnarray*}
&&
\sum_{i\in\mathcal{P}_k}c_k\biggl\{\biggl(\alpha_{i1}^{(k)}-\alpha_i-b_k\sum_{j\in A_M}w_j^{(k)}w_{ij}^*\biggr)^2-\bigl(\alpha_{i1}^{(k)}-\alpha_i\bigr)^2\biggr\}
\\
&&\quad{}+
\sum_{i\in\mathcal{P}_k}\sum_{t\in\mathcal{P}_k^c}c_k\biggl[\biggl\{\alpha_{t1}^{(k)}
-
\alpha_t+b_k\sum_{j\in A_M}w_j^{(k)}d_{ij}\Delta_{jk}w_{tj}^*\biggr\}^2-\bigl(\alpha_{t1}^{(k)}-\alpha_t\bigr)^2\biggr]
\\
&&\qquad=
\sum_{i\in\mathcal{P}_k}\{\alpha_i^2-\alpha_i-\phi_i\},
\end{eqnarray*}
which generalizes (\ref{14}). Here, we assume common variances for the
units in the same PSU.

We extend the fractional nearest neighbor imputation to the case of
$M_1$ fractions for point estimation and $M_2$ (${>}M_1$) fractions for
variance estimation. The motivation for this extension is the
application to the Census long form where the official estimates are
based on a single imputed value. A~second imputed value was generated
to be used only in variance estimation.
Let~$d_{1ij}$ and $d_{2ij}$ be the
donor--recipient relationship indicator function used for point
estimation and for variance estimation, respectively.\vspace*{1pt} Also, let
$w_{1ij}^*$ and~$w_{2ij}^*$ be the fractional weights of recipient $j$
from donor $i$ that are computed from $d_{1ij}$ and $d_{2ij}$,
respectively. For missing unit $j$, one common choice is $w_{1ij}^* =
d_{1ij} M_1^{-1}$ and $w_{2ij}^* = d_{2ij} M_2^{-1}$. Of particular
interest is the case where $M_1 = 1$ and $M_2 = 2$.

If $M_1 \neq M_2$, the variance estimator is defined by
\begin{equation}\label{9c}
\hat{V} ( \hat{\theta}_I ) = \sum_{k=1}^L c_k \bigl( \hat{\theta
}_I^{(k)} -
\hat{\theta}_I \bigr)^2 ,
\end{equation}
where
\[
\bigl( \hat{\theta}_I^{(k)}, \hat{\theta}_I \bigr) = \biggl( \sum_{ i \in A_R }
\alpha_{i2}^{(k)} y_i ,\sum_{ i \in A_R } \alpha_{i1} y_i \biggr)
\]
with $\alpha_{i2}^{(k)} =\sum_{j } w_j^{(k)} w_{2ij}^{*(k)} $ and
$\alpha_{i1} =\sum_{j } w_j w_{1ij}^* $. Here,
$w_{2ij}^{*(k)}$ is the
replicated fractional weight of unit $j$ assigned to donor $i$ in the
$k$th replication. Note that $\hat{\theta}_I$ is based on the point
estimation weights and $\alpha_{i2}^{(k)}$ is based on the variance
estimation weights.
If $w_{2ij}^{*(k)}$ satisfy (\ref{con1}),
the bias of the variance estimator (\ref{9c}) is
\[
\mathit{Bias} \{\hat{V} \}
= E\Biggl\{ \sum_{i \in A_R } \Biggl[ \sum_{k=1}^L c_k \bigl( \alpha_{i2}^{(k)} -
\alpha_{i1} \bigr)^2-( \alpha_{i1}^2 -\alpha_{i1} ) \Biggr] \sigma_i^2 \Biggr\}.
\]
Thus, condition (\ref{con2}) for the unbiasedness of the variance
estimator is changed to
\begin{equation}\label{con2b}
\sum_{k=1}^L\! c_k \biggl\{\! \bigl( \alpha_{i2}^{(k)} \!-\! \alpha_{i1} \bigr)^2 \!+\!\! \sum_{ t
\in D_{Ri} } \!\bigl( \alpha_{t2}^{(k)} \!-\! \alpha_{t1} \bigr)^2 \!\biggr\} = \alpha_{i1}^2
\!-\! \alpha_{i1} \!+ \!\!\sum_{t \in D_{Ri} }\! ( \alpha_{t1}^2 \!-\! \alpha_{t1}
).\hspace*{-30pt}
\end{equation}

To create the replicated fractional weights satisfying (\ref{con1}) and
(\ref{con2b}), the sum of squares of the naive replication weights is
first computed,
\[
\sum_{k=1}^L c_k \bigl( \alpha_{i1}^{(k)} - \alpha_{i1} \bigr)^2 = \phi_{i1},\qquad i
\in A_R,
\]
where $\alpha_{i1}^{(k)}=\sum_{j \in A} w_j^{(k)}
w_{1ij}^{*}$.
In the second step the
fractions for replicates for donors in the point estimation are
modified.
Let
the new fractional weight in replicate $k$ for the value donated by $i
\in\mathcal{P}_k$ to $j$ be
\[
w_{2ij}^{*(k)} = w_{1ij}^*( 1 - b_k) , \qquad\mbox{if } i \in\mathcal{P}_k
\mbox{ and } M_2 \neq M_{2jk},
\]
where $b_k$ is to be determined and $M_{2jk} = \sum_{ i \in
\mathcal{P}_k } d_{2ij} $. Now, $M_2$ (${>}M_1$) donors are identified for
variance estimation. The new fractional weight for the other $M_2-1$
donors to recipient $j$, denoted by $t$, is
\begin{equation}\label{13c}
w_{2tj}^{*(k)} = w_{1tj}^* + \Delta_{jk} b_k d_{1ij} w_{2tj}^* \qquad\mbox{for } t \in\mathcal{P}_k^c \mbox{ and } i \in\mathcal{P}_k,
\end{equation}
where
\[
\Delta_{jk} =
\frac{ \sum_{i \in\mathcal{P}_k } w_{1ij}^* }{ \sum_{i \in
\mathcal{P}_k^c } w_{2ij}^* }.
\]
Then the $b_k$ that gives the correct sum
of squares is the solution to the quadratic equation
\begin{eqnarray*}
&&
\sum_{ i \in\mathcal{P}_k } c_k \biggl\{ \biggl( \alpha_{i1}^{(k)} -
\alpha_{i1} - b_k \sum_{ j \in A_M} w_j^{(k)} w_{1ij}^* \biggr)^2 - \bigl(
\alpha_{i1}^{(k)} - \alpha_{i1} \bigr)^2 \biggr\}
\\
&&\quad{}+
\sum_{ i \in\mathcal{P}_k } \sum_{t \in\mathcal{P}_k^c}c_k \biggl[
\biggl\{ \alpha_{t1}^{(k)} - \alpha_{t1} + b_k \sum_{j \in A_M} w_j^{(k)}
\Delta_{jk}d_{1ij} w_{2tj}^* \biggr\}^2 - \bigl( \alpha_{t1}^{(k)} - \alpha_{t1}\bigr)^2\biggr]
\\
&&\qquad=
\sum_{ i \in\mathcal{P}_k }( \alpha_{1i}^2 - \alpha_{1i} -
\phi_{1i} ).
\end{eqnarray*}

If $M_1=1$, the adjustment in the replication fractional weights can be
made at the individual level. Let the new fractional weight in
replicate $k$ for the value donated by $i \in\mathcal{P}_k$ to $j$, $j
\in\mathcal{P}_k^c$, be
\[
w_{2ij}^{*(k)} = w_{1ij}^*( 1 - b_i) , \qquad\mbox{if } i \in\mathcal{P}_k
\mbox{ and } M_2 \neq M_{2jk},
\]
where $b_i$ is to be determined. The new fractional weight for each of
the other $M_2-1$ donors to recipient $j$, denoted by $t$, is
\[
w_{2tj}^{*(k)} = w_{1tj}^* + \Delta_{jk} b_i d_{1ij} w_{2tj}^* \qquad\mbox{for } t \in\mathcal{P}_k^c \mbox{ and } i \in\mathcal{P}_k,
\]
where $ \Delta_{jk} $ is defined following (\ref{13c}).
Then the $b_i$ that gives the correct sum
of squares is the solution to the quadratic equation
\begin{eqnarray*}
&&
c_k \biggl\{ \biggl( \alpha_{i1}^{(k)} - \alpha_{i1} - b_i \sum_{ j \in A_M}
w_j^{(k)} w_{1ij}^* \biggr)^2 - \bigl( \alpha_{i1}^{(k)} - \alpha_{i1} \bigr)^2 \biggr\}
\\
&&\quad{}+
 \sum_{t \in\mathcal{P}_k^c}c_k \biggl[ \biggl\{ \alpha_{t1}^{(k)} -
\alpha_{t1} + b_i \sum_{j \in A_M} w_j^{(k)} \Delta_{jk}d_{1ij}
w_{2tj}^* \biggr\}^2 - \bigl( \alpha_{t1}^{(k)} - \alpha_{t1} \bigr)^2
\biggr]
\\
&&\qquad=
\alpha_{1i}^2 - \alpha_{1i} - \phi_{1i} .
\end{eqnarray*}

\section{Application to US Census long form data}\label{s5}

\subsection{Introduction}
We use long form data from the states of Delaware and Michigan to
provide examples of the variance estimation methods.
Table \ref{table1} shows
the individual income items and their state level imputation rates for
Delaware and Michigan.

\begin{table}
\tablewidth=300pt
\caption{Imputation rate and the person-level average income for each
income item (age${}\ge 15$) for two states, Delaware ($n=87{,}280$) and
Michigan ($n=1{,}412{,}339$)} \label{table1}
\begin{tabular*}{\tablewidth}{@{\extracolsep{\fill}}lcccc@{}}
\hline
& \multicolumn{2}{c}{\textbf{Delaware}} & \multicolumn{2}{c@{}}{\textbf{Michigan}} \\[-5pt]
& \multicolumn{2}{c}{\hrulefill} & \multicolumn{2}{c@{}}{\hrulefill}\\
 & \textbf{Imputation} & \textbf{Average} & \textbf{Imputation} & \textbf{Average} \\
\textbf{Income item}& \textbf{rate (\%)} & \textbf{income} & \textbf{rate (\%)} & \textbf{income} \\
\hline
Wage & 20 & 21,892 & 21 & 20,438 \\
Self employment & 10 & \phantom{0,}1286 & 10 & \phantom{0,}1234\\
Interest & 22 & \phantom{0,}1989 & 22 & \phantom{0,}1569\\
Social security & 20 & \phantom{0,}1768 & 20 & \phantom{0,}1672 \\
Supplemental security & 20 & \phantom{00,}125 & 20 & \phantom{00,}148\\
Public assistance & 19 & \phantom{000,}38 & 19 & \phantom{000,}47 \\
Retirement & 20 & \phantom{0,}2018& 20 & \phantom{0,}1664\\
Other & 19 & \phantom{00,}543 & 19 & \phantom{00,}529\\[3pt]
Total & 31 & 29,659& 31 & 27,301 \\
\hline
\end{tabular*}
\end{table}

The sampling design for the Census 2000 long form used stratified
systematic sampling of households, with four strata in each state.
Sampling rates varied from 1 in 2 for very small counties and small
places to 1 in 8 for very populous areas.

The weighting procedure for the Census 2000 long form was performed
separately for person estimates and for housing unit estimates. For the
income and poverty estimates considered here, the person weights are
needed.

The census long form person weights are created in two steps. In the
first step, the initial weights are computed as the ratio of the
population size (obtained from the 100\% population counts) to the
sample size in each cell of a~cross-classification of final weighting
areas (FWAs) by person types [Housing unit person, Service Based
Enumeration (SBE) person, other Group Quarters (GQ) person]. Thus, the~initial weights take the form of post-stratification weights. The
second step in the weighting is raking, where, for person weights,
there are four dimensions in the raking. The dimensions are household
type and size (21 categories), sampling type (3 categories),
householder classification (2 categories), and Hispanic
origin/race/sex/age (312 categories). Therefore, the total number of
possible cells is 39,312, although many cells in a FWA will be empty.
The raking procedure is performed within each FWA. There are about
60,000 FWAs in the whole country and the FWAs are nested within
counties.

\subsection{Computational details}

The variance estimation methodology is ba\-sed on the grouped jackknife,
where the method described in Section \ref{s3} is used to estimate the
variance due to imputation. We summarize the main steps of variance
estimation and then discuss the steps in more detail:
\begin{description}
\item[Step 1:] Create groups and then define initial replication
weights for the grouped jackknife method. The elements within a stratum
are systematically divided into groups. A replicate is created by
deleting a group.

\item[Step 2:] Using the initial replication weights, repeat the
weighting procedure to compute the final weights for each replicate.

\item[Step 3:] Using fractional weighting, modify the replicate
weights to account for the imputation effect on the variance. In the
process, a replicate imputed total income variable is created for each
person with missing data.

\item[Step 4:] Using the replicate total income variables, compute
the jackknife variance estimates for parameters such as the number of
poor people by age group and the median household income.
\end{description}

In step 1, the sample households in a final weighting area are sorted
by their identification numbers, called MAFIDs. Let $n$ be the sample
number of households in a final weighting area. The first $n/50$ sample
households are assigned to variance stratum 1, the next $n/50$ sample
households are assigned to variance stratum 2, and so on, to create 50
variance strata. Within each variance stratum, the sample households
are further grouped into two groups by a systematic sample of
households arranged in a half-ascending-half-descending order based on
the MAFID. Using the two groups in each of the 50 strata, $L=100$
replication factors are assigned to each unit in the sample. For unit
$i$ in variance stratum $h$ $( h=1,2,\ldots,50) $, the replication
factor for the replicate formed by deleting group~$k$ in variance
stratum~$h$ is
\[
F_{i}^{(hk)}=\cases{
1, &\quad if unit $i$ does not belong to variance stratum $h$, \cr
2-\delta_{i}, &\quad if unit $i$ belongs to variance stratum $h$ and $i\notin\mathcal{P}_{hk}$, \cr
\delta_{i}, &\quad if unit $i\in\mathcal{P}_{hk}$,%
}
\]
where $\delta_{i}=1-\{ ( 1-1/w_{i0}) 0.5\} ^{1/2}$, $%
w_{i0}$ is the initial weight of unit $i$, and~$\mathcal{P}_{hk}$ is
the set of sample indices in group $k$ in variance stratum $h$. With
this replication factor, $c_{k}$ of (\ref{9}) is one.

In step 2, the step 1 replication weights are modified using the
production raking operation. The weighting procedure consists of two
parts. The first part is a poststratification in each final weighting
area and the second part is raking ratio estimation using the short
form population totals as controls. If the raking was carried to
convergence, the estimated variance for controls would be zero. In the
actual operation, the replicated final weights produce very small
variance estimates for the estimates of the population controls.

In step 3, a second nearest neighbor is identified for each
nonrespondent for each income item. There are eight income items---see
Table \ref{table1} given earlier. A fractional weight of one is assigned to the
imputed value from the first donor and a fractional weight of zero is
assigned to the imputed value from the second donor for production
estimation. The fractional weights are changed for the replicate, when
the jackknife group containing the first donor is deleted. The amount
of change is determined so that conditions (\ref{con1}) and
(\ref{con2}) are satisfied. Replicate fractional weights are
constructed separately for each income item.

Once the replicated fractional weights are computed, replicates of the
person-level total income are constructed. Let $Y_{tis}$ be the $s$th
income item for person $i$ in family $t$ and let $R_{tis}$ be the
response indicator function for~$Y_{tis}$. For the $k$th replicate, the
replicated total income for person~$i$ in fami\-ly~$t$ is\looseness=1
\begin{equation}\label{ptotr}
\mathit{TINC}_{ti}^{(k)}=\sum_{s=1}^{8}\bigl\{ R_{tis}Y_{tis}+( 1-R_{tis})
Y_{tis}^{\ast(k)}\bigr\},
\end{equation}
where $Y_{tis}^{\ast(k)}$ is the $k$th replicate of the imputed value
for $%
Y_{tis}$, defined by
\[
Y_{tis}^{\ast(k)}=w_{tisa}^{\ast(k)}Y_{tisa}^{\ast}+w_{tisb}^{\ast
(k)}Y_{tisb}^{\ast},
\]
$( w_{tisa}^{\ast(k)},w_{tisb}^{\ast(k)}) $ is the vector of the two
$k$th replicate fractional weights, one for the first donor and one for
the second donor, for the $s$th income item, and $( Y_{tisa}^{\ast
},Y_{tisb}^{\ast}) $ is the vector of the imputed values of $Y_{tis}$
from the first and second donor, respectively. The $k$th replicate of
total family income for family $t$ is
\begin{equation}\label{htotr}
\mathit{TINC}_{t}^{(k)}=\sum_{i=1}^{m_{t}}\mathit{TINC}_{ti}^{(k)},
\end{equation}
where $m_{t}$ is the number of people in family $t$ and
$\mathit{TINC}_{ti}^{(k)}$ is defined in~(\ref{ptotr}).

For the age group poverty estimates, a poverty status indicator
function is defined for the family, and applies to all family members.
That is, all family members are either in poverty or all are not in
poverty. The poverty status indicator for family $t$ is defined as
\[
\zeta_{t}=\cases{
1, &\quad if $\mathit{TINC}_{t}<c_{t}$, \cr
0, &\quad if $\mathit{TINC}_{t}\geq c_{t}$,%
}
\]
where, as with the replicates in (\ref{ptotr}),
\[
\mathit{TINC}_{t}=\sum_{i=1}^{m_{t}}\sum_{s=1}^{8}\{ R_{tis}Y_{tis}+( 1-R_{tis})
Y_{tisa}^{\ast}\}
\]
is the total income of family $t$, where $Y_{tisa}^{\ast}$ is the
imputed value for $Y_{tis}$ using the first nearest donor, and $c_{t}$
is the poverty threshold value for family $t$. The threshold is a
function of the number of related children under 18 years of age, the
size of the family unit, and the age of the householder. (Poverty
thresholds for all recent years are available on the Census Bureau web
site at
\url{http://www.census.gov/hhes/www/poverty/threshld.html}.)\vadjust{\eject}

To compute the replicate of $\zeta_t$, we use the following procedure:
\begin{enumerate}[2.]
\item For person $i$ in family $t$, compute two total incomes,
$\mathit{TINC}_{tia} $ and $\mathit{TINC}_{tib}$, by
\begin{eqnarray*}
\mathit{TINC}_{tia}
&=&
\sum_{s=1}^{8}\{ R_{tis}Y_{tis}+( 1-R_{tis})Y_{tisa}^{\ast}\},
\\
\mathit{TINC}_{tib}
&=&
\sum_{s=1}^{8}\{ R_{tis}Y_{tis}+( 1-R_{tis})Y_{tisb}^{\ast}\} .
\end{eqnarray*}
Also, compute the two total family incomes
\[
( \mathit{TINC}_{ta},\mathit{TINC}_{tb}) =\sum_{i=1}^{m_{t}}( \mathit{TINC}_{tia},\mathit{TINC}_{tib}) .
\]
Using the replicated total family income $\mathit{TINC}_{t}^{(k)}$ defined in
(\ref{htotr}), define
\begin{equation}\label{all}
\alpha_{t}^{(k)}=\frac{\mathit{TINC}_{t}^{(k)}-\mathit{TINC}_{tb}}{\mathit{TINC}_{ta}-\mathit{TINC}_{tb}}, \qquad \mbox{if }\mathit{TINC}_{ta}\neq\mathit{TINC}_{tb},
\end{equation}
and
$\alpha_{t}^{(k)}=1$ otherwise.
The $\alpha_{t}^{(k)}$ is the weight satisfying
\[
\mathit{TINC}_{t}^{(k)}=\alpha_{t}^{(k)}\mathit{TINC}_{ta}+\bigl( 1-\alpha_{t}^{(k)}\bigr)
\mathit{TINC}_{tb}.
\]
\item The replicated poverty status variable is now computed by
\begin{equation}\label{38}
\zeta_{t}^{(k)}=\alpha_{t}^{(k)}\mathit{POV}_{ta}+\bigl( 1-\alpha_{t}^{(k)}\bigr)
\mathit{POV}_{tb},
\end{equation}
where $\mathit{POV}_{ta}$ is computed by
\[
\mathit{POV}_{ta}=\cases{
1, &\quad if $\mathit{TINC}_{ta}<c_{t}$, \cr
0, &\quad if $\mathit{TINC}_{ta}\geq c_{t}$
}
\]
and $\mathit{POV}_{tb}$ is computed similarly using $\mathit{TINC}_{tib}$.
\end{enumerate}
The replication adjustment $\alpha_{t}^{(k)}$ is computed
from family-level total income and is applied in (\ref{38}) to get a
replicated poverty estimate.

The estimated variance for the estimated total number of people in
poverty is
\begin{equation}\label{var}
\hat{V}_p =\sum_{k=1}^{L} \bigl( \hat{\theta}_p^{(k)} - {\hat{\theta}}
_p^{(\cdot)} \bigr)^2,
\end{equation}
where $L$ is the number of replications (here $L=100$),
\[
\hat{\theta}_p^{(k)} = \sum_{t=1}^n \sum_{i=1}^{m_t} w_{tj}^{(k)}
\zeta_t^{(k)},
\qquad
{\hat{\theta}}_p^{(\cdot)} = \frac{1}{L} \sum_{k=1}^{L} \hat
{\theta}
^{(k)}_p,
\]
$\zeta_t^{(k)}$ is defined in (\ref{38}), and $w_{ti}^{(k)}$ is the
person level replication weight after the raking operation.

The number of people in poverty in a given age group can be estimated~%
by
\[
\hat{\theta}_{pz}=\sum_{t=1}^{n}\sum_{i=1}^{m_{t}}w_{ti}z_{ti}\zeta
_{t},
\]
where $z_{ti}=1$ if the person $i$ in family $t$ belongs to the age
group and $z_{ti}=0$ otherwise. The $k$th replicate of the estimate is
\[
\hat{\theta}_{pz}^{(k)}=\sum_{t=1}^{n}\sum
_{i=1}^{m_{t}}w_{ti}^{(k)}z_{ti}%
\zeta_{t}^{(k)}
\]
and the variance is estimated by (\ref{var}) using
$\hat{\theta}_{pz}^{(k)}$ defined above.

The variance estimation for median household income estimates is based
on the test-inversion methodology described in Francisco and Fuller
(\citeyear{ff1991}). Also, see Woodruff (\citeyear{w1952}). Let $\mathit{MED}$ be the estimated median
household income defined by $\mathit{MED}=\hat{F}^{-1}( 0.5) $, where $\hat{F}(
\cdot) $ is the estimated cumulative distribution function of total
income of the household,
\[
\hat{F}( u) =\Biggl( \sum_{t=1}^{n}w_{tt}\Biggr) ^{-1}\sum_{t=1}^{n}w_{tt}I(
\mathit{TINC}_{t}\leq u) ,
\]
$w_{tt}$ is the householder's person weight in household $t$, and
$\mathit{TINC}_{t}$ is the total income of household $t$. (Note that households
differ from families. The former includes all persons living in a given
housing unit; the latter includes only related persons living in a
housing unit.)

To apply the test-inversion method, first create the replicated
indicator variable
\[
\mathit{INV}_{t}^{(k)}=\alpha_{t}^{(k)}\mathit{INV}_{ta}+\bigl( 1-\alpha_{t}^{(k)}\bigr)
\mathit{INV}_{tb},
\]
where $\alpha_{t}^{(k)}$ is defined in (\ref{all}) and
\[
\mathit{INV}_{ta}=\cases{
1, &\quad if $\displaystyle\sum_{i=1}^{m_{t}}\mathit{TINC}_{tia}<\mathit{MED}$, \cr
0, &\quad if $\displaystyle\sum_{i=1}^{m_{t}}\mathit{TINC}_{tia}\geq \mathit{MED}$
}
\]
and $\mathit{INV}_{tb}$ is computed similarly, using $\mathit{TINC}_{tib}$ instead of $%
\mathit{TINC}_{tia}$ in the above expressions.

The estimated variance of the estimated proportion $\hat{F} ( \mathit{MED}
)=0.5$ is computed by applying the variance formula (\ref{var}) using $
\mathit{INV}_t^{(k)}$ instead of~$\zeta_t^{(k)}$ to get $\hat{V}_{\mathrm{inv}}$. Define
\[
( \hat{p}_1 , \hat{p}_2 ) = \bigl( 0.5 - 2 \sqrt{\hat{V}_{\mathrm{inv}}}, 0.5 +
2\sqrt{\hat{V}_{\mathrm{inv}}} \bigr)
\]
to be an approximate 95\% confidence interval for the estimated
proportion $%
\hat{F} ( \mathit{MED} )=0.5$. The estimated variance of the estimated median is
\[
\hat{V}_{\mathrm{med}} = \{ \hat{F}^{-1} ( \hat{p}_2 ) -\hat{F}^{-1} ( \hat{p}_1
) \}^2/16 .
\]

\subsection{Numerical results}

Variance estimates for the long form income and poverty estimates that
have been used by SAIPE were computed for all 50 states of the US (plus
DC) and their counties. The estimates considered here are the total
number of people in poverty, the number of children under age~5 in
poverty (state level only), the number of related children age 5--17
in families in poverty, the number of children under age 18 in poverty,
and the median household income.

\begin{table}
\tablewidth=330pt
\caption{Variance estimation results for Delaware and Michigan}\label{table2}%
\vspace*{-3pt}
\begin{tabular*}{\tablewidth}{@{\extracolsep{\fill}}lccccc@{}}
\hline
 & & \multicolumn{2}{c}{\textbf{Delaware}} & \multicolumn{2}{c@{}}{\textbf{Michigan}}\\[-5pt]
&  & \multicolumn{2}{c}{\hrulefill} & \multicolumn{2}{c@{}}{\hrulefill} \\
\textbf{Parameter}& \textbf{Method }& \textbf{Est. SE} & \textbf{Std. SE} & \textbf{Est. SE} & \textbf{Std. SE} \\
\hline
$\theta_1$ & Naive & \phantom{0}870 & 100 & 3217 & 100\\
(total in poverty) & Imputation & 1161 & 133 & 4096 & 127 \\
$\theta_2$ & Naive & \phantom{0}221 & 100& \phantom{0}776 & 100\\
(0--4 in poverty) & Imputation & \phantom{0}260 & 118 & \phantom{0}897 & 116 \\
$\theta_3$ & Naive & \phantom{0}366 & 100& 1314 & 100 \\
(5--17 related in poverty) & Imputation & \phantom{0}467 & 128& 1640 & 125\\
$\theta_4$ & Naive & \phantom{0}458 & 100 & 1608 & 100 \\
(0--17 in poverty) & Imputation & \phantom{0}592 & 129& 2062 & 128 \\
Median & Naive & \phantom{0}177 & 100& \phantom{00}70 & 100\\
HH income & Imputation & \phantom{0}207 & 117 & \phantom{00}85 & 121\\
\hline
\end{tabular*}
\vspace*{-5pt}
\end{table}

Table \ref{table2} contains variance estimation results (the estimated
standard deviations) for the income and poverty statistics for the
states of Delaware and Michigan. The variance estimator labeled
``naive'' treats the imputed values as observed values. The
``imputation'' variance estimator is that of Section \ref{s3} and reflects the
imputation effects. Both variance estimators account for the raking in
the estimator. Because Michigan is much larger than Delaware, its
estimated numbers of persons in poverty (not shown) are much larger,
and thus, due to the scale effects, so are the corresponding standard
errors. The standardized standard errors in the table are computed by
dividing the estimated standard error computed by the ``imputation''
procedure by the estimated standard error computed by the ``naive''
procedure.

\begin{table}
\tablewidth=200pt
\caption{ Imputation rates by income level (age ${}\ge 15$)}\label{table3}%
\vspace*{-3pt}
\begin{tabular*}{\tablewidth}{@{\extracolsep{\fill}}lcc@{}}
\hline
 & \multicolumn{2}{c@{}}{\textbf{Imputation rate (\%)}} \\[-5pt]
& \multicolumn{2}{c@{}}{\hrulefill} \\
\textbf{Total income}& \textbf{Delaware} & \textbf{Michigan} \\
\hline
\phantom{10,00}0--9999 & 34 & 34 \\
10,000--19,999 & 36 & 35 \\
20,000--49,999 & 28 & 29 \\
50,000--69,999 & 25 & 25 \\
70,000 and over & 25 & 25 \\
\hline
\end{tabular*}
\vspace*{-5pt}
\end{table}

Generally speaking, imputation increases the variance so the naive
variance estimator underestimates the true variance. The relative
increase is similar for Michigan and Delaware. A result worth noting is
that the increase in variance due to imputation is higher for the
poverty parameters than for the income parameters. This is because in
both states the imputation rate is higher for persons with low imputed
income. (See Table \ref{table3}.)

\begin{table}[b]
\tablewidth=300pt
\caption{County variance estimates for Delaware }\label{table4}%
\vspace*{-3pt}
\begin{tabular*}{\tablewidth}{@{\extracolsep{\fill}}lcccc@{}}
\hline
\textbf{County} & \textbf{Parameter} & \textbf{Method} & \textbf{Est. SE} & \textbf{Std. SE} \\
\hline
001 & $\theta_1$ & Naive & 409 & 100 \\
& (total poor) & Imputation & 444 & 109 \\
& $\theta_3$ & Naive & 183 & 100\\
& (5--17 related poor ) & Imputation & 203 & 111 \\
&$\theta_4$ & Naive & 219 & 100 \\
& (0--17 poor) & Imputation & 241 & 110 \\
& Median & Naive & 323 & 100\\
& HH income & Imputation & 336 & 104\\[3pt]
003 & $\theta_1$ & Naive & 687 & 100 \\
& (total poor) & Imputation & 838 & 122 \\
& $\theta_3$ & Naive & 317 & 100\\
& (5--17 related poor) & Imputation & 351 & 111 \\
&$\theta_4$ & Naive & 365 & 100 \\
& (0--17 poor) & Imputation & 417 & 114 \\
& Median & Naive & 200 & 100\\
& HH income & Imputation & 226 & 113\\[3pt]
005 & $\theta_1$ & Naive & 518 & 100 \\
& (total poor) & Imputation & 608 & 117 \\
& $\theta_3$ & Naive & 197 & 100\\
& (5--17 related poor) & Imputation & 217 & 110 \\
&$\theta_4$ & Naive & 270 & 100 \\
& (0--17 poor) & Imputation & 300 & 111 \\
& Median & Naive & 361 & 100\\
& HH income & Imputation & 389 & 108\\
\hline
\end{tabular*}
\vspace*{-5pt}
\end{table}

Table \ref{table4} contains some numerical results for the estimated
standard errors for the county estimates in Delaware. The age groups in
the table are those used by SAIPE at the county level, which are fewer
than the age groups used by SAIPE at the state level. As with state
estimates, imputation increases the variance. However, the effect of
imputation is much smaller for county estimates than for state
estimates. County level estimation is an example of domain estimation,
where the values used for imputation can come from donors outside the
domain. Donors from outside the domain contribute less to the
imputation variance of the domain total than donors in the domain
because the imputed value from outside the domain is uncorrelated with
the values observed in the domain. In effect, imputations from outside
the domain increase the sample size on which the estimates are based,
whereas imputations from inside the domain change the weights given to
the observations in the estimates. Because the proportions of outside
donors differ across counties, the effect of imputation on county
variances is not uniform across counties. In Delaware, the overall
imputation rates for total income (the percent of records with at least
one income item imputed) are 30.7\%, 29.5\%, and 34.5\% for county 1,
county 3, and county 5, respectively. Table \ref{table5} presents the
distribution of donors for wage income in Delaware. In~coun\-ty~1, about
$59\%$ of the donors are from outside the county, whereas in~coun\-ty~3,
only about $25\%$ of the donors are from outside the county. Thus, the
variance inflation due to imputation, as reflected by the standardized
standard error, is greater for county 3 than for county 1.

\begin{table}
\tablewidth=310pt
\caption{Donor distribution for wage income in Delaware (age${}\ge
15$)}\label{table5}
\begin{tabular*}{\tablewidth}{@{\extracolsep{\fill}}lccc@{}}
\hline
 & \textbf{Number of donors} &\textbf{ Number of donors} & \textbf{Number of donors} \\
\textbf{County}& \textbf{from county 1} & \textbf{from county 3} & \textbf{from county 5} \\
\hline
1 & 1271 & 1512 & 325 \\
$(n=15{,}735)$ & (41\%) & (49\%) & (10\%) \\
3 & 1142 & 7374 & 1343 \\
$(n=51{,}869)$ & (11\%) & (75\%) & (14\%) \\
5 & 847 & 1137 & 2045 \\
$(n=19{,}661)$ & (21\%) & (28\%) & (51\%) \\
\hline
\end{tabular*}
\end{table}

\section*{Acknowledgments}
We thank two anonymous referees and the Associate Editor for very
helpful comments. The research was supported by a contract with the US
Census Bureau.
We also
thank George McLaughlin and George Train for computational support and
Yves Thibaudeau for discussion on the long form imputation methods.

\begin{supplement}[id-suppA]
\sname{Supplement A}
\stitle{Illustrated calculations}
\slink[doi]{10.1214/10-AOAS419SUPPA}
\slink[url]{http://lib.stat.cmu.edu/aoas/419/supplement_a.pdf}
\sdatatype{.pdf}
\sdescription{We illustrate the construction of replicates for
variance estimation with a simple example where a simple random sample
of original size six is selected with two missing values and two donors
per missing value.}
\end{supplement}

\begin{supplement}[id-suppA]
\sname{Supplement B}
\stitle{Justification for (\ref{1})}
\slink[doi]{10.1214/10-AOAS419SUPPB}
\slink[url]{http://lib.stat.cmu.edu/aoas/419/supplement_b.pdf}
\sdatatype{.pdf}
\sdescription{We provide a justification for (\ref{1})
based on the large sample theory. The assumptions and the proof for
(\ref{1}) are provided.}
\end{supplement}

\begin{supplement}[id-suppA]
\sname{Supplement C}
\stitle{Proofs}
\slink[doi]{10.1214/10-AOAS419SUPPC}
\slink[url]{http://lib.stat.cmu.edu/aoas/419/supplement_c.pdf}
\sdatatype{.pdf}
\sdescription{Proofs for equations (\ref{8}), (\ref{10}), and
(\ref{6b}) are provided.}
\end{supplement}


\printaddresses

\end{document}